\documentclass[fleqn,usenatbib]{mnras}

\usepackage{newtxtext,newtxmath}

\usepackage[T1]{fontenc}

\DeclareRobustCommand{\VAN}[3]{#2}
\let\VANthebibliography\thebibliography
\def\thebibliography{\DeclareRobustCommand{\VAN}[3]{##3}\VANthebibliography}

\usepackage{graphicx}	
\usepackage{subcaption}
\usepackage{amsmath}	
\usepackage{float}
\usepackage{stfloats}
\usepackage[a4paper]{geometry}
\usepackage{multicol}

\title[Statistics of Pressure Fluctuations in Turbulent Kinetic Plasmas]{Statistics of Pressure Fluctuations in Turbulent Kinetic Plasmas}

\author[S. Adhikari et al.]{
S. Adhikari,$^{1,2}$\thanks{E-mail: subash.adhikari@mail.wvu.edu}
W. H. Matthaeus,$^{2,3}$
T. N. Parashar,$^{2,4}$
M.A. Shay,$^{2,3}$
and P. A. Cassak$^{1,2,3}$
\\
$^{1}$ Department of Physics and Astronomy, West Virginia University, Morgantown, West Virginia 26506, USA\\
$^{2}$ Department of Physics and Astronomy,
University of Delaware,
Newark, Delaware 19716, USA\\
$^{3}$ Bartol Research Institute, Department of Physics and Astronomy,University of Delaware, Newark, Delaware 19716, USA\\
$^{4}$ School of Chemical and Physical Sciences, Victoria University of Wellington, Wellington 6140, NZ
}

\date{Accepted XXX. Received YYY; in original form ZZZ}

\pubyear{2023}

\begin{document}
\label{firstpage}
\pagerange{\pageref{firstpage}--\pageref{lastpage}}
\maketitle

\begin{abstract}
In this study we explore the statistics of pressure fluctuations in kinetic collisionless turbulence. A $2.5$D kinetic particle-in-cell (PIC) simulation of decaying turbulence is used to investigate pressure balance via the evolution of thermal and magnetic pressure in a plasma with $\beta$ of order unity. We also discuss the behavior of thermal, magnetic and total pressure structure functions and their corresponding wavenumber spectra. The total pressure spectrum exhibits a slope of $-7/3$ extending for about a decade in the ion-inertial range. In contrast, shallower $-5/3$ spectra are 
characteristic of the 
magnetic pressure and thermal pressure. The steeper total pressure spectrum is a consequence of cancellation caused by density-magnetic field magnitude anticorrelation. Further, we evaluate higher order total pressure structure functions in an effort to discuss intermittency and compare the power exponents with higher order structure functions of velocity and magnetic fluctuations. Finally, applications to astrophysical systems are also discussed.
\end{abstract}




\section{Introduction}
\label{sec:intro}
\subsection{Motivation}

Most naturally occurring as well as laboratory plasma flows are turbulent. Astrophysical systems of interest include the intracluster medium (ICM)~\citep{schuecker2004probing, churazov2012x, zhuravleva2014turbulent, zhuravleva2019suppressed}, black hole accretion disks~\citep{balbus1998instability, pessah2010angular}, solar wind~\citep{coleman1968turbulence, matthaeus1982measurement, bruno2013solar}, and planetary magnetospheres~\citep{sahraoui2020magnetohydrodynamic}. Turbulent stresses and pressure play an important role at scales ranging from the largest structures in the universe~\citep{simionescu2019constraining} to laboratory plasmas. 
A prominent feature of galactic plasma observations is the ``great powerlaw''  spectrum of electron density fluctuations that 
extends over more than ten orders of magnitude in scale~\citep{armstrong1981density}
in the interstellar medium. 
Under mild assumptions~\citep{montgomery1987density} this spectrum is
found to be consistent with  Kolmogorov spectrum of 
magnetic field fluctuations.
This calculation depends on an assumption of near-incompressibility and also implies that density and magnetic field, and therefore mechanical and magnetic pressures, are anticorrelated and partially cancel one another. 
This kind of {\it statistical pressure balance}
is an element of turbulent dynamics, in contrast to the 
superficially similar 
{\it static pressure balances}
that are special solutions of 
the ideal MHD equations. 

Pressure balance plays an important role in astrophysics. The mass of galaxy clusters ~\citep{KravtsovARAA12, planelles2015large}, the largest structures in the universe,
provides stringent constraints on cosmological models~\citep{voigt2006galaxy, carlberg1997average, pratt2019galaxy}. Hydrostatic pressure balance is a common assumption used to estimate the mass of relaxed clusters \citep{lau2013weighing}. However, this assumption can easily break in clusters far from equilibrium and can also be broken by turbulent stress contributions~\citep{markevitch2004direct,angelinelli2020turbulent}. X-ray~\citep{schuecker2004probing, churazov2012x, zhuravleva2014turbulent} and radio observations via the Sunayev-Zel'dovich~\citep{khatri2016thermal} effect provide detailed measurements of the thermal properties of galaxy clusters. An accurate understanding of the pressure dynamics in the intracluster medium can help disentangle contributions from thermal pressure and turbulent stresses.

Similarly, properties of the 
pressure have a major impact in the heliosphere.
Near-Earth, the solar wind plays a vital role in shaping Earth's magnetosphere and driving space weather. The solar wind is measured to be turbulent~\citep{coleman1968turbulence, matthaeus1982measurement, bruno2013solar} and is observed to be hotter than expected from adiabatic cooling~\citep{wang2001energy, hellinger2013proton}. 
Heating implies an increase of the trace of
the pressure tensors, and this increase in the internal energy 
can occur through 
interaction between the pressure tensors
and the velocity gradient tensors~\citep{yang2017energy}. The solar wind routinely displays intervals of pressure balance ~\citep{vellante1987analysis, goldstein1995magnetohydrodynamic, reisenfeld1999evidence, ruffolo2021domains} which potentially emerge as a result of the nearly incompressible magnetohydrodynamics (NI-MHD) of the solar wind~\citep{matthaeus1991nearly}. 

In this study, we develop a framework to discuss pressure balance
in the context of 
the equation governing pressure in incompressible MHD with and without shear. We also explore various aspects of pressure dynamics in kinetic plasmas using a fully kinetic particle-in-cell (PIC) simulation. We find the pressure spectrum follows the hydrodynamic scaling in the incompressible MHD regime. Although the kinetic plasmas are not fully incompressible, the pressure spectrum shows an excellent agreement with the power scaling following the MHD predictions, indicating that the simulation is likely in a state of NI-MHD. The remaining paper is organized as follows: 
Section \ref{sec:theoretical} first
reviews
some additional theoretical background and then constructs the Poisson's equation for pressure in an incompressible MHD framework. In section \ref{sec:results} we report the results from PIC simulation, and finally we discuss the implications and future directions in section \ref{sec:conc}.

\section{Theory}\label{sec:theoretical}
\subsection{Background}\label{sec:background}
Following the seminal work on 
hydrodynamic 
pressure fluctuations by Obukhoff and Batchelor~\citep{obukhoff1951microstructure,batchelor1951pressure},the second-order pressure structure function $D_P^2$ and pressure spectrum have been studied both experimentally ~\citep{uberoi1953quadruple,george1984pressure,tsuji2003similarity} and via direct numerical simulations (DNS) ~\citep{schumann1978numerical,kim1993isotropy,pullin1994pressure,nelkin1998scaling, squire2023pressure}, mostly for homogeneous and isotropic turbulence. While these studies rely on the assumptions that the velocity distribution is joint gaussian or the small-scale statistics follow Kolmogorov's similarity hypothesis, $D_P^2$ as a function of lag $r$ is found to be proportional to $r^{4/3}$. Therefore, the pressure spectrum $E_P(k)$ varies as $k^{-7/3}$ within the inertial range, where $k$ is the wavenumber. Attempts have also been made to extend the pressure spectrum to the higher wavenumber (dissipation range)~\citep{zhao2016pressure}. However, 
the kinetic/dissipation range scaling exponents 
are less well understood, in general and 
for the pressure spectrum. 

The study of nearly incompressible
MHD~\citep{KlainermanMajda81} 
is a bit more subtle than the hydro case. Beginning with the \cite{montgomery1987density} explanation of the $k^{-5/3}$ spectrum of the interstellar density structure function,
several additional
layers of theory have been presented~\citep{Higdon84, MattBrown88}
based on the idea that the core solution of the plasma dynamics
in a perturbation theory controlled by the turbulent Mach number is incompressible
MHD, and the leading order density 
fluctations are those discussed by \citet{montgomery1987density}.
Additional theories have been presented that elaborate on 
the role of thermal fluctuations ~\citep{ZankMatt90, BaylyEA92}, 
and variations in plasma beta~\citep{ZankMatt92b, ZankMatt93}.
In light of the latter references, the original perturbation 
theory of \cite{montgomery1987density} and \cite{MattBrown88}
can be seen to apply most readily to plasma beta greater than unity. 

To place the current work better in 
the context of these prior studies, 
it is important to emphasize that 
 most of the literature focuses on mechanical pressure 
 statistics in both 
 hydrodynamic turbulence and MHD. 
 Then mechanical pressure is linked to density and thermal fluctuations through an equation of state. 
 In contrast, the present study is mainly concerned with 
statistics of the total pressure -- magnetic plus mechanical -- and contrasting this with behavior of the magnetic and mechanical pressure separately. Therefore we are concerned with pressure balances. 
We will also examine higher 
order statistics related to total pressure. 
Notably we will study in detail these properties
using a particle-in-cell model of 
collisonless plasma, thus departing substantially from the starting 
point of the MHD model. 

\subsection{Pressure Poisson's equation in MHD}\label{sec:theory}
The incompressible MHD equations (using Einstein's convention) are given by
\begin{eqnarray}
\partial_t u_i+ u_j\partial_j u_i - b_j\partial_j b_i=- \partial_i P +\nu \partial_j \partial_j u_i,\\
\partial_t b_i +u_j\partial_j b_i -b_j \partial_j u_i =\eta \partial_j \partial_j b_i,
\label{MHDb}
\end{eqnarray}
where $\textbf{u}$ is the flow velocity, $\textbf{b}$ is the magnetic field in Alfv\'en units, $P$ is the total pressure (magnetic plus thermal), $\nu$ is the kinematic viscosity, and $\eta$ is the electric diffusivity. Here, the density is assumed to be unity
and the velocity is solenoidal $\nabla \cdot {\bf u} = 0$. 
Now, in terms of Els\"{a}sser variables $\textbf{z}^\pm$ defined as $\textbf{z}^\pm = \textbf{u} \pm \textbf{b}$, the above equations can be combined as
\begin{equation}
    \partial_t  z_i^\pm+ z_j^\mp \partial_j z_i^\pm = -\partial_i P + \nu \partial_j \partial_j z_i^\pm,
\label{MHDz}
\end{equation}
where we assume $\nu=\eta$. 
Taking the divergence of Eqn.~\ref{MHDz} and 
employing 
incompressibility ($\partial_i z_i=0$) we get
\begin{equation}
     -\partial_{ii} P=\partial_i(z_j^\mp\partial_j z_i^\pm)=(\partial_i z_j^\mp)(\partial_j z_i^\pm).
\label{MHDpress}
\end{equation}
As written, the pressure is now seen to 
be as a constraint force that acts to maintain the solenoidal nature of the velocity field. The so-called
{\it Nearly Incompressible MHD} (NI-MHD)
theory proceeds to compute the density as a linear response to the pressure fluctuations \citep{batchelor1951pressure,montgomery1987density,matthaeus1988nearly} that emerge from the solution of Eqn.~\ref{MHDpress}.

Incompressibility does not imply constant pressure in the system, as seen in Eqn.~\ref{MHDpress}; instead, pressure is a non-linear function of the Els\"{a}sser variables suggesting any fluctuation in the Els\"{a}sser variable creates a pressure field in the system and vice-versa. The right hand side of Eqn.~\ref{MHDpress} can be split into symmetric and anti-symmetric (skew-symmetric) parts as
\begin{equation}
     -\partial_{ii} P=(\partial_i z_j^\mp) (\partial_j z_i^\pm) =(\tau_{ij}^\mp+R_{ij}^\mp)(\tau_{ij}^\pm-R_{ij}^\pm),
\label{MHDexpansion}
\end{equation}
where $\tau_{ij}^\pm=\frac{1}{2}(\partial_i z_j^\pm +\partial_j z_i^\pm)$ is the symmetric part of the decomposition, and is called the rate of strain tensor related to the Els\"{a}sser's variable, while $R_{ij}^\pm=\frac{1}{2}(\partial_i z_j^\pm -\partial_j z_i^\pm)$ is the anti-symmetric component, called the rate of rotation tensor associated with the curl of the Els\"{a}sser variable. One can simplify Eqn.~\ref{MHDexpansion} further to obtain
\begin{equation}
     -\partial_{ii} P=\tau_{ij}^-\tau_{ij}^+ -R_{ij}^-R_{ij}^+.
\label{MHDsimplify}
\end{equation}

The first term on the right hand side is related to the strain tensor. If one decomposes $\tau_{ij}^\pm$ such that $S_{ij}=\frac{1}{2}(\partial_i u_j +\partial_j u_i)$ and $M_{ij}=\frac{1}{2}(\partial_i b_j +\partial_j b_i)$ are the velocity and magnetic rates of strain respectively, then $\tau_{ij}^-\tau_{ij}^+= S_{ij}S_{ij}-M_{ij}M_{ij}=S^2-M^2$, where $S^2$ and $M^2$ are the respective squares of the 
velocity and magnetic rate of strain tensors.
These are the associated second tensor invariants, equal to the sum of squares of the respective eigenvalues. 

The second term on the right hand side simplifies further to yield $R_{ij}^-R_{ij}^+=\frac{1}{4} (\nabla \times \mathbf{z}^-)_k (\nabla \times \mathbf{z}^+)_k=\frac{1}{4}(\omega_k-j_k)(\omega_k+j_k)=\frac{1}{4}(\omega_k\omega_k-j_kj_k)$, where $\mathbf{\omega}=\nabla \times \textbf{u}$ is the vorticity and $\mathbf{j}=\nabla \times \textbf{b}$ is the current density. Therefore, Eqn.~\ref{MHDsimplify} can be re-written as
\begin{equation}
    -\partial_{ii} P=(S^2-M^2)-\frac{1}{4}(\omega_i\omega_i-j_ij_i).
\label{eqn:pressurefinal}
\end{equation}
Eqn.~\ref{eqn:pressurefinal} is the Poisson equation for pressure, which shows that the pressure fluctuations are related to the rate of velocity and magnetic field strains and their rotations. 
In the absence of a magnetic field, Eqn.~\ref{eqn:pressurefinal} reduces to $-\partial_{ii} P=S^2-\frac{1}{4}\omega_i\omega_i$.

\subsection{Poisson's equation with shear}\label{sec:poisson}
Using Reynolds' decomposition, we can write a physical quantity (say $P$) as the sum of mean ($\overline{P}$) and fluctuating component ($\tilde{p}$). Substituting in Eqn.~\ref{MHDpress}, and using $\partial_i z_i^\pm=0$, we get
\begin{equation}
     \partial_{ii} \overline{P}+\partial_{ii} \tilde{p}=-\partial_i \partial_j(\overline{Z}_i^\mp\overline{Z}_j^\pm+\overline{Z}_i^\mp\tilde{z}_j^\pm+\tilde{z}_i^\mp\overline{Z}_j^\pm+\tilde{z}_i^\mp\tilde{z}_j^\pm).
\label{sheareqnint}
\end{equation}
Subtracting the ensemble average from itself we get
\begin{equation}
     \partial_{ii} \tilde{p}=-\partial_i \partial_j(\overline{Z}_i^\mp\tilde{z}_j^\pm+\tilde{z}_i^\mp\overline{Z}_j^\pm) - \partial_i \partial_j(\tilde{z}_i^\mp\tilde{z}_j^\pm-\overline{\tilde{z}_i^\mp \tilde{z}_j^\pm}),
\label{sheareqnfinal}
\end{equation}
which is the shear form of the Poisson equation for fluctuations in pressure.

The form of Eqn.~\ref{sheareqnfinal} is the 
same as its hydrodynamic counterpart 
except that $u$ is replaced by $z$. 
Clearly, the Laplacian of pressure fluctuation is composed of two different terms. The first term represents the distortion produced by the shear as a result of interaction of shear with turbulence, while the second term is the interaction among the turbulent fluctuations. Solving Eqns.~\ref{eqn:pressurefinal} and ~\ref{sheareqnfinal} for pressure requires specific boundary conditions which is out of scope of this paper. Instead, we focus ourselves on the statistics of pressure, as follows. 
The characteristics of pressure fluctuation can be studied using the pressure correlation functions $ R_P(\mathbf{r}) =\langle P(\mathbf{x}+\mathbf{r})P(\mathbf{x})\rangle$ ~\citep{batchelor1951pressure} or the second-order pressure structure functions $D^2_P(\textbf{r})=\lvert \delta P(\textbf{r}) \rvert^2$ ~\citep{obukhoff1951microstructure} both of which are related to each other~\citep{monin1975statistical}. Here, $\textbf{x}$ represents the positions space, $\mathbf{r}$ represents the spatial lag,  $\langle...\rangle$ represents the ensemble average, and the pressure increment is defined as $\delta P (\textbf{r}) = P(\textbf{x}+\textbf{r})-P(\textbf{x})$. 

The scaling of the pressure spectrum in the inertial range can be readily identified by applying Kolmogorov similarity hypothesis to pressure increments \citep{kolmogorov1991local}. The second-order structure function $D^2_P(r)$ depends only on spatial lag $r$ and rate of dissipation of energy $\epsilon$, and can be written as
\begin{equation}
    D^2_P(r) =\langle|P(\mathbf{x}+\mathbf{r})-P(\mathbf{x})|^2\rangle = C_p \epsilon ^{4/3} r^{4/3},
\label{stfeqn}
\end{equation}
where $C_p$ is a constant. The form in Eqn.~\ref{stfeqn} is valid in the inertial range. A Fourier transfrom of Eqn.~\ref{stfeqn} results in the pressure spectrum $E_P(k)$ with the mathematical form
\begin{equation}
    E_P(k)\propto \epsilon ^{4/3} k^{-7/3},
    \label{eqn:pressurespectrum}
\end{equation}
where $k$ is the wavenumber. The second-order pressure structure function has also been linked to the fourth-order velocity structure function ~\citep{hill1995pressure, hill1997pressure, nelkin1998scaling} and pressure-gradient velocity-velocity structure function ~\citep{hill2001next}. However, exploring such relationship is beyond the scope of this paper, where our emphasis is on kinetic plasmas.

\section{Simulation and Results}
\label{sec:results}
To study the statistics of pressure fluctuations in kinetic plasmas, we analyze a $2.5$D fully kinetic particle-in-cell (PIC) simulation of turbulence performed using the P3D code~\citep{zeiler2002three}. The simulation follows the normalization where length is normalized to the ion-inertial length $d_i=c/\omega_{pi}$, where $c$ is the speed of light and $\omega_{pi}$ is the plasma frequency for ions; time is normalized to the inverse of the cyclotron frequency $\omega_{ci}$ and velocity is normalized to Alfv\'en speed $v_{A0}=B_0/\sqrt{4\pi m_i n_0}$, where $m_i$ is the mass of ions, $B_0$ and $n_0$ are the normalizing parameters for magnetic field and number density, respectively.
Similarly, temperature is normalized to $m_i v_{A0}^2$ and pressure is normalized to $n_0 m_i v_{A0}^2$. 

The simulation is a periodic square domain of length $L=149.6d_i$ with grid points of $4096^2$, and $3200$ particles initially in each grid. The mass ratio of ions to electrons is set to $25$ with an initial background density of $1$, uniform initial temperature of $T_e=T_i=0.3$ with a total plasma beta $\beta=1.2$, and an initial out-of-plane magnetic field of strength $1$. The system is initially populated with Fourier modes for magnetic field and velocity (both ions and electrons) within $2k_0\leq|k|\leq 4k_0$, where $k_0=2\pi/L$ and allowed to evolve for about $370\, t\omega_{ci}$ without external forcing
(see ~\cite{parashar2018dependence},~\cite{adhikari2021energy} for details).

\subsection{Pressure balances}\label{sec:balance}
\begin{figure}
    \centering
    \includegraphics[scale=0.7]{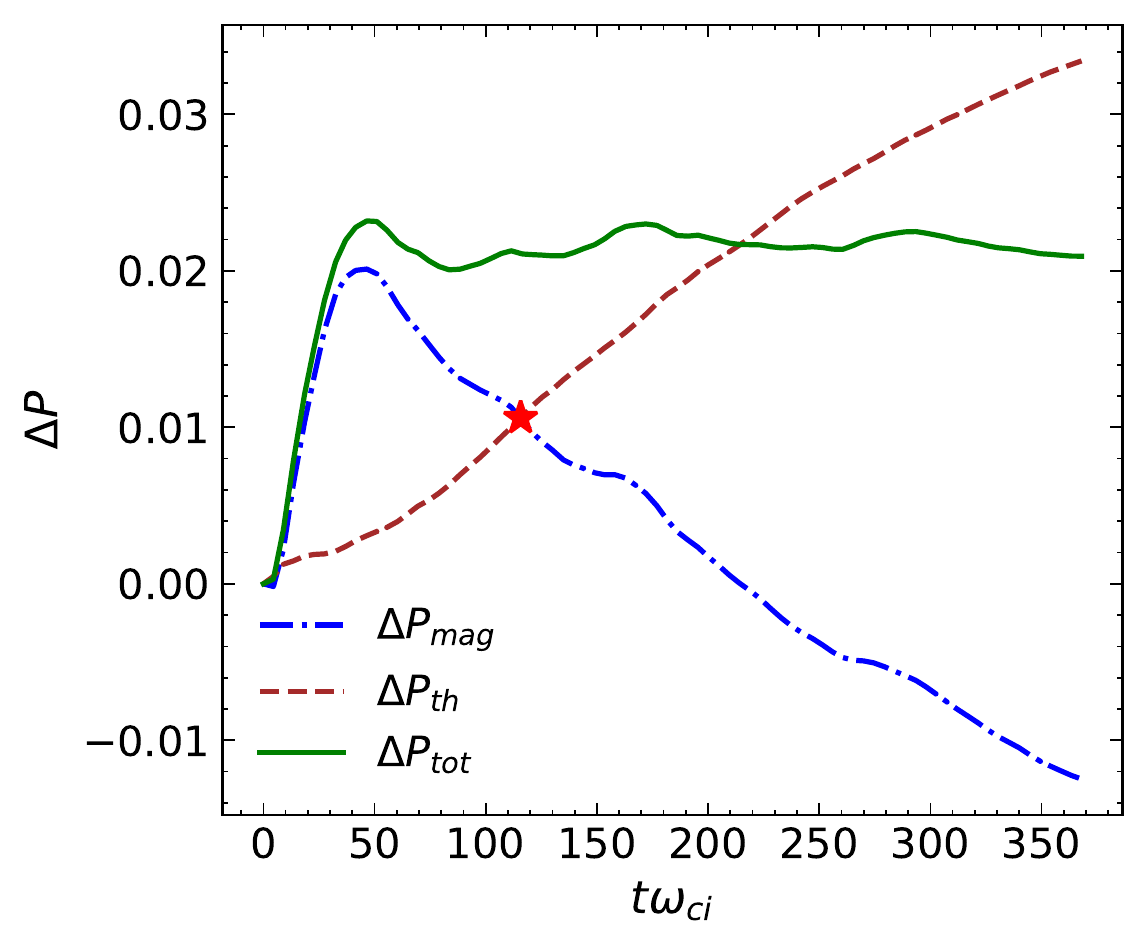}
    \caption{\label{fig:deltaP} Time evolution of the change in thermal $P_{th}$, magnetic $P_{mag}$ and total pressure $P_{tot}$ calculated with respect to their initial values. $P_{mag}$ and $P_{th}$ have opposite behavior once the magnetic pressure reaches the maximum value. The red star represents the time of analysis.}
\end{figure}

Fig. \ref{fig:deltaP} shows the time evolution of the ensemble averaged (i.e., volume integrated) 
change in magnetic $\Delta P_{mag}$ and total thermal $\Delta P_{th}$ pressure. The change in the pressure is calculated with respect to the initial value at $t=0$. Initially, the magnetic pressure increases rapidly for a small period of time and then falls, while the thermal pressure continues to increase gradually throughout the simulation. Note, the change in the total pressure stays roughly constant once it reaches the maximum value; as the thermal pressure compensates for the fall in the magnetic pressure. The change in the magnetic and thermal pressure intersects at $t\omega_{ci}=116.5$ when the mean square current is maximum in the system (not shown).
This time, when MHD dissipation typically reaches a maximum, is often regarded as the turbulence having reached a fully developed state. One finds that in this 
state, the volume-integrated magnetic pressure and the volume-integrated thermal pressure are anticorrelated, producing a near-constant volume integrated total pressure. 
Since we are interested in the turbulence properties of pressure fluctuations, we focus our analysis around this time. 

\begin{figure*}
\centering
\includegraphics[scale=0.9]{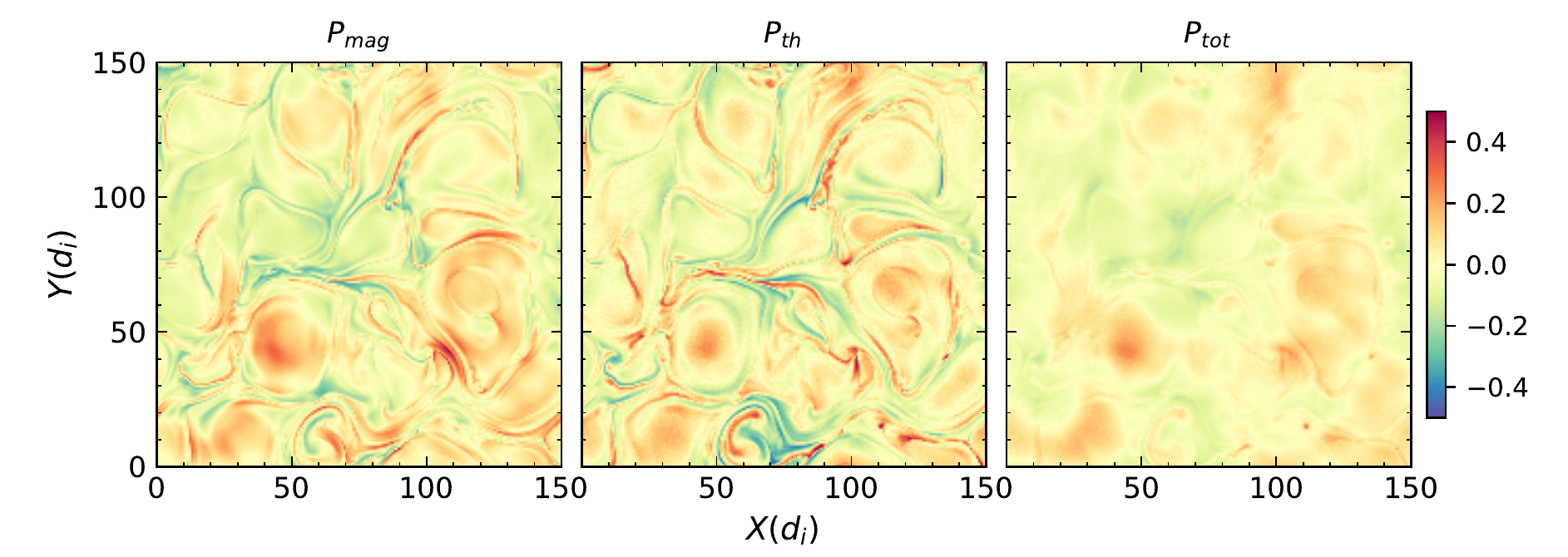}
\caption{Two dimensional image  at $t\omega_{ci}=116.5$ of the magnetic $P_{mag}$, thermal $P_{th}$, and total $P_{tot}$ pressure fluctuations normalized to their respective mean values.}
\label{fig:pressure2D}
\end{figure*}

\begin{figure}
\centering
\includegraphics[scale=0.65]{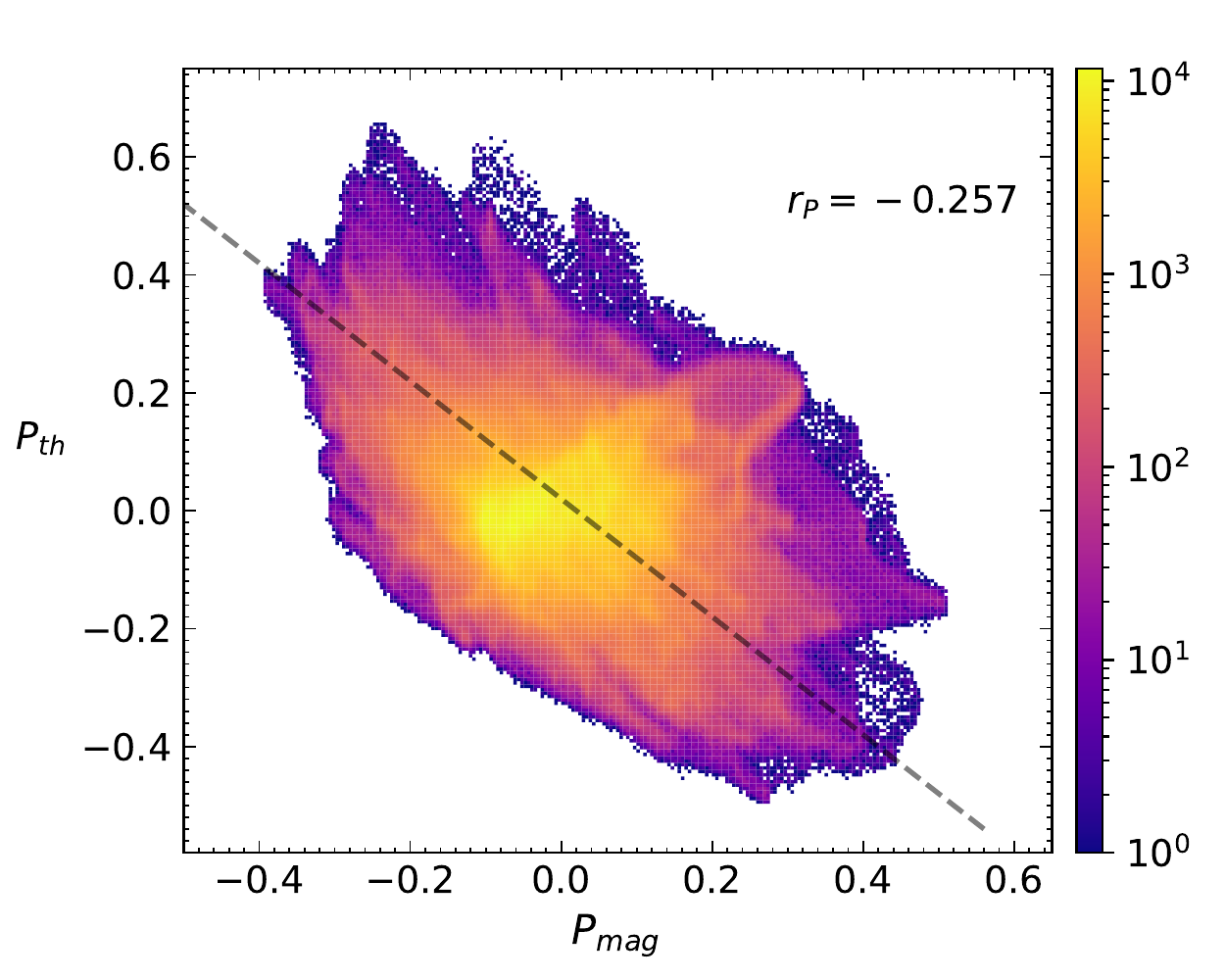}
\caption{Joint probability distribution of the thermal $P_{th}$ and magnetic $P_{mag}$ pressure at $t\omega_{ci}=116.5$. A dashed line of slope $-1$ is drawn for reference and $r_P$ represents the Pearson correlation coefficient.}
\label{fig:joint_pdf}
\end{figure}

To understand how the global pressure balance 
impacts the local pressures, 
in Fig. \ref{fig:pressure2D} we show the two-dimensional view of the thermal, magnetic, and total pressure fluctuations at $t\omega_{ci}=116.5$, each normalized to their respective mean value. It is observed that most of the regions near the current sheets (see Fig.\ref{fig:jzmag}) have low magnetic pressure and high thermal pressure indicating a negative correlations between the two. As a result the total pressure in these regions stay relatively uniform as seen on the right panel of Fig.~\ref{fig:pressure2D}. On the other hand, the magnetic islands have similar magnitude of $P_{mag}$ and $P_{th}$. This is certainly due to the magnetic compression and heating within the islands. The total pressure balance is further quantified by calculating the variance of the pressure fluctuations plotted in Fig.~\ref{fig:pressure2D}, and is shown in Table.~\ref{tab:variance}. The variance of the magnetic and thermal pressure fluctuations have similar magnitude, while the variance of the total pressure fluctuation is much smaller ($\sim 40 \%$ of $P_{th}$ or $P_{mag}$).

Next, we plot the joint probability distribution function (PDF) of the thermal and magnetic pressure in Fig.~\ref{fig:joint_pdf}, where a line of slope $-1$ is drawn for reference. The joint PDF clearly follows the reference line suggesting a negative correlation between the two pressures. In addition, we compute the Pearson correlation coefficient $r_P$~\citep{cohen2009pearson} between $P_{th}$ and $P_{mag}$ to be $-0.257$ supporting the trend seen in Fig.~\ref{fig:joint_pdf}. This anticorrelation quantifies the tendency towards statistical (in contrast 
to pointwise) pressure balance. 

\begin{table}
\centering
\caption{Variance about the spatial mean of the magnetic, thermal and total pressure as plotted in Fig.~\ref{fig:pressure2D}.}
\label{tab:variance}
\begin{tabular}{l|c|c|r} 
\hline
Variance & $P_{th}$ & $P_{mag}$ & $P_{tot}$\\
\hline
$\sigma^2$ & $0.0107$ & $0.0101$ & $0.0039$\\
\hline
\end{tabular}
\end{table}

\begin{figure}
\centering
\includegraphics[scale=0.6]{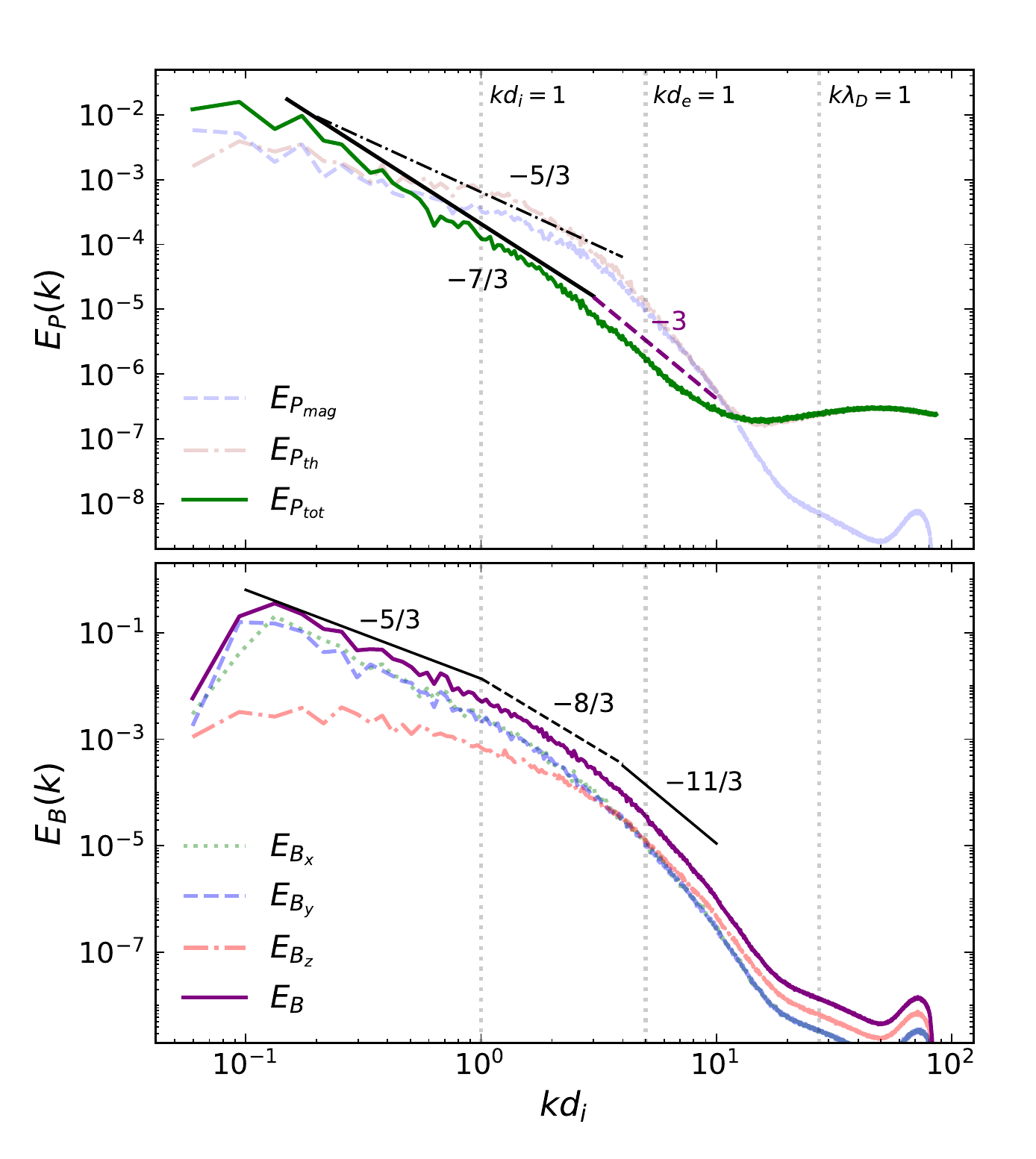}
\caption{(Top) Pressure spectra for magnetic
pressure (dashed line), 
thermal pressure (dash-dot)
and total pressure (solid green line)
as a function of wavenumber at the time of analysis. Reference lines with 
slopes $-5/3$, $-7/3$ and $-3$ are 
provided. 
(Bottom) Magnetic energy spectra at the same time. Lines of slope $-5/3$, $-8/3$, and $-11/3$ are drawn for reference. Both total spectrum and component spectra are shown (See legend). }
\label{fig:pstf_spectra}
\end{figure}

We now discuss the pressure spectra.
In the top panel of Fig.~\ref{fig:pstf_spectra}, we plot the magnetic, thermal, and total pressure spectra. The dotted vertical lines represent the wavenumber corresponding to the ion inertial length $d_i^{-1}$ , electron inertial length $d_e^{-1}$ and the Debye length $\lambda_D^{-1}$. Additional lines of slope $-5/3$, $-7/3$, and $-3$ are drawn for reference. The thermal and magnetic pressure spectra exhibit very similar behavior except at the higher wavenumbers. In the inertial range $kd_i \lesssim 1$, both these spectra exhibit similar spectral slopes close to $-5/3$, while at large wavenumbers $kd_i >5$, the thermal pressure has a plateau at approximately $10^{-7}$ while the magnetic pressure spectrum falls down below $10^{-8}$.

Once again, the total pressure displays different behavior compared to its constituent parts. The total pressure spectrum, in the inertial range displays a spectral slope of $-7/3$ consistent with Eqn.~\ref{eqn:pressurespectrum}. This is due to the significant cross-spectrum correlation between the thermal and the magnetic pressure, a direct consequence of the statistical pressure balance described above. 

At higher wavenumber, the total pressure spectrum steepens beyond the inertial range $2 \leq kd_i \leq 9$ with slope $-3$ before the hump on the thermal pressure spectrum dominates the contribution. 

The bottom panel of Fig.~\ref{fig:pstf_spectra} shows the magnetic energy spectra along with its components. 
In the inertial range 
($kd_i \lesssim 1$) 
the magnetic spectra 
 approximate the familiar Kolmogorov $-5/3$ slope as expected in strong turbulence.
Near $kd_i =1$ the magnetic spectra steepen, as often reported for plasma 
\citep{LeamonEA98-jgr}, with a slope that varies in the range of about $-7/3$ to $-8/3$ \citep{SmithEA06-diss}.
In this range kinetic dissipation and dispersion effects become important. 
Various steeper forms have been proposed for the spectrum at still higher wavenumbers $kd_e>1$; an $11/3$ spectrum is shown 
here for reference. 
The spectrum of the out of plane component $B_z$
is much flatter than the other two magnetic components, and in the inertial range appears at a much lower amplitude, down by as much as two orders of magnitude at the long wavelength end of the inertial range. However
at the intermediate kinetic range, the $B_z$ spectrum reaches equipartition with the spectra of the other two components, and further along, at sub electron scales, this component becomes dominant over the other two by a modest factor. 


We note in passing that 
the theoretical development in \citep{montgomery1987density}
began with assumption that the magnetic spectrum exhibits a $-5/3$ spectrum
and proceeded to compute the density and pressure spectra as a linear response to the 
magnetic field. The conclusion was that in MHD the thermal pressure 
(and density) spectra
would take on a $-5/3$ power law as seen here. The anticorrelation of 
thermal and magnetic pressure at inertial range scales was also concluded. 
The statistical 
characterization of total pressure was not included. The present findings are complementary to this important antecedent. 

\begin{figure}
    \includegraphics[scale=0.5]{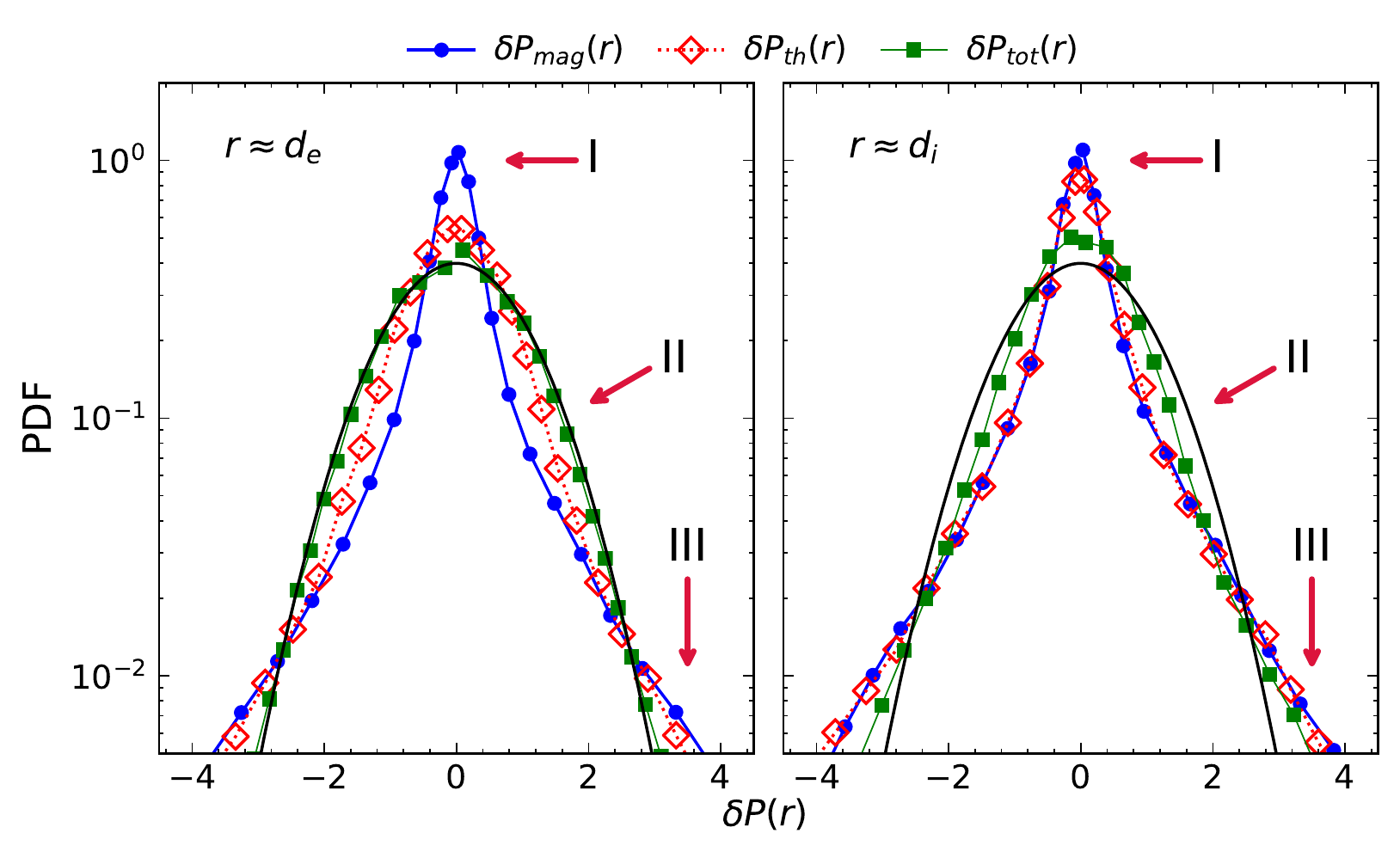}
     \caption{Probability density functions of pressure increments $\delta P$, for
     magnetic pressure (blue circles),  
     thermal pressure (red diamonds) and
     total pressure (green boxes),
     each normalized by its standard deviation, 
     at $r\approx 1 d_e =0.2 d_i$ (left), and $r \approx 1d_i$ (right). The PDFs are computed 
     at $t=116.5\omega_{ci}^{-1}$. The black solid line 
     is a normal distribution for reference (see text for details).}
   \label{fig:pressurepdf}
\end{figure}

Next, we show the probability density function (PDF) of the pressure increments $(\delta P)$ defined in section~\ref{sec:poisson} as
\begin{equation}\label{eqn:pressure_increment}
    \delta P\mathbf{(r})= P(\mathbf{x}+\mathbf{r})-P(\mathbf{x}),
\end{equation}
where $\mathbf{r}$ is the spatial lag. For simplicity, we compute the PDF using lags along the $x$ and $y$ directions, and average the results. One can also perform these calculations for lags along different directions and take an angular average~\citep{wang2022strategies}. The PDF's are shown in Fig. \ref{fig:pressurepdf} for two different lags $r=1d_e \approx 0.2 d_i$ (left panel) and $r\approx 1 d_i$ (right panel). 

At $r \approx d_i$, the PDF of both $\delta P_{mag}$ (blue circles) and $\delta P_{th}$ (red diamonds) follow each other closely in contrast to the $r\approx d_e$ case, however, displaying a departure from gaussianity in both cases. The total pressure PDF (green box) for $r\approx d_e$ is very close to a normal distribution (solid black line), drawn for reference, but is distorted for $r\approx d_i$ implying better pressure balance at smaller lags. 

\begin{figure*}
    \includegraphics[scale=0.6]{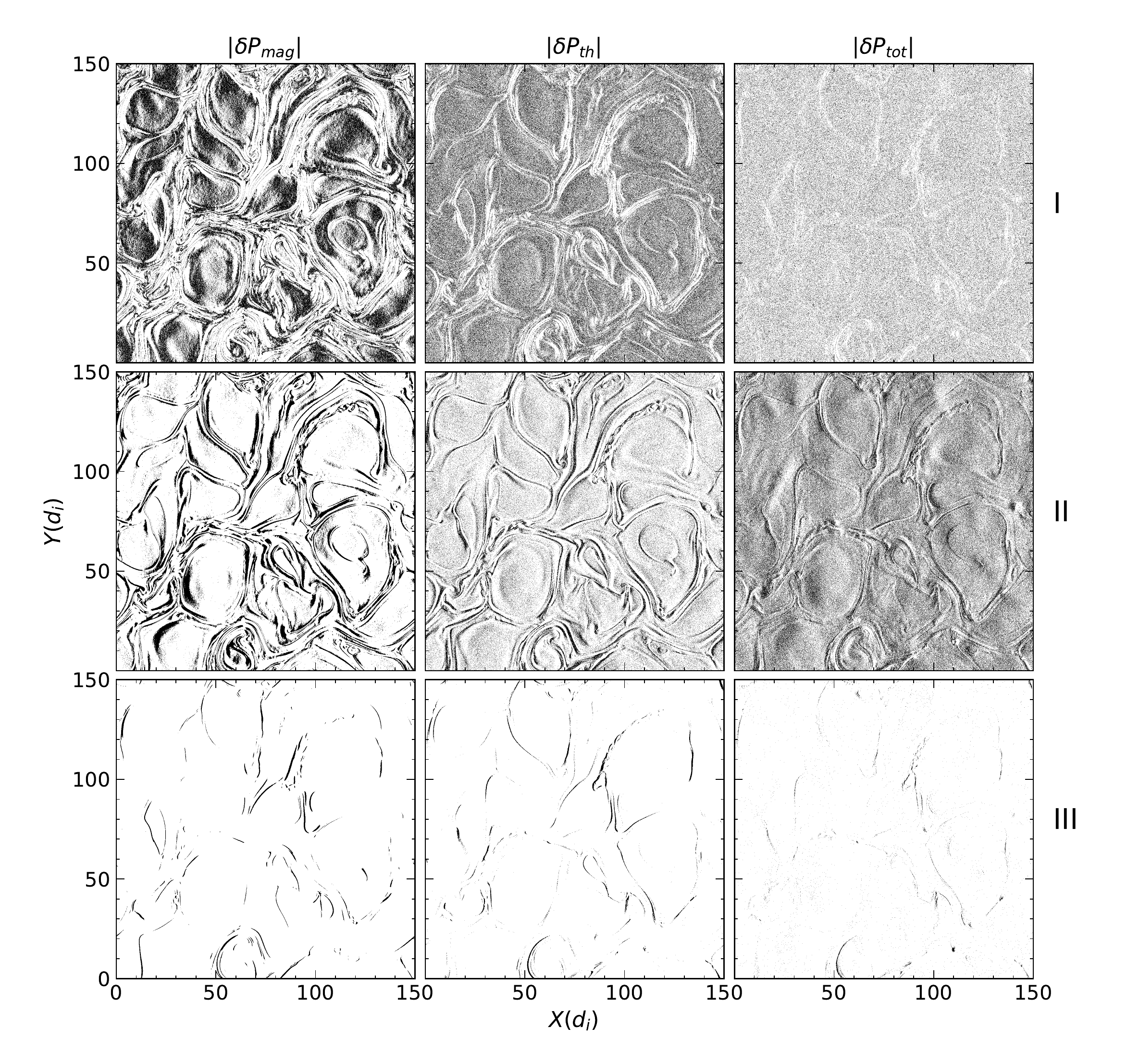}
     \caption{Structures (in black) contributing to regions I, II and III of the probability density function (PDF) of the magnetic (left column), thermal (middle column) and total pressure (right column) increments shown in Fig.~\ref{fig:pressurepdf} for a lag of $r\approx d_e$, averaged over $x$ and $y$ directions. Each row corresponds to the region shown on the right side of the figure. }
   \label{fig:pdffilterall}
\end{figure*}

To examine the spatial distributions that originate 
from different sections of these 
PDF's, we divide the probability distribution function into regions I, II and III (as in ~\cite{GrecoEA09-ace}) and mask out all but one region in a given plot. The results for
small lag increments, at $r\approx d_e$, are shown in Fig.~\ref{fig:pdffilterall}. Here we have used a binary color representation where black represents the structures contributing to a particular region, and white represents the region with no contribution.
It is immediately evident that the total pressure increments at this scale are much more homogeneous
than the separate contributions from magnetic or thermal 
pressures. This is the case for the region I core of the distributions, for the fine scale super-gaussian tails representing region III, and for the 
intermediate subgaussian regions II. 
This signature of pressure balance is particularly 
clear for the supergaussian tail. 
However none of these regions is completely uniform. 
In fact the spatial distribution of magnetic islands and their boundaries, i.e., what we sometimes called 
cellularization of the plasma~\citep{servidio2008depression}, is evident in all 
three ranges of the PDF, 
as is evident in 
the panels of Fig.~\ref{fig:pdffilterall}.
Quantitative diagnostics of
this spatial structure leads us 
to a discussion of {\it intermittency}. 

Theories and classifications of 
intermittency provide a physical basis for 
quantifying deviations from gaussianity. 
To understand the intermittency of pressure fluctuations, we calculate the scale-dependent kurtosis and higher-order structure function for pressure increments in the following sub-section.


\subsection{Higher Order Statistics and Intermittency}
\subsubsection{Scale dependent kurtosis}
The scale-dependent kurtosis $\kappa(\mathbf{r})$ for the pressure fluctuations $\delta P(\mathbf{r})$ as a function of lag $\mathbf{r}$ is defined as
\begin{equation}
    \kappa(\mathbf{r})=\frac{\langle\lvert P(\mathbf{x}+\mathbf{r}) -P(\mathbf{x}) \rvert^4\rangle}{\langle \lvert P(\mathbf{x}+\mathbf{r}) -P(\mathbf{x}) \rvert^2\rangle^2}.
\label{eqn:sdk}
\end{equation}
$\kappa(\mathbf{r})$ takes on a value of three for a scalar field having a gaussian distribution. 
For a scalar quantity (or its increments) 
less randomly distributed and more clustered in space, $\kappa(\mathbf{r})$ attains larger values, and can be heuristically interpreted as the reciprocal of the spatial filling factor. This property makes $\kappa(\mathbf{r})$ a good preliminary indicator of the presence of spatial intermittency. 

\begin{figure}
\centering
\includegraphics[scale=0.52]{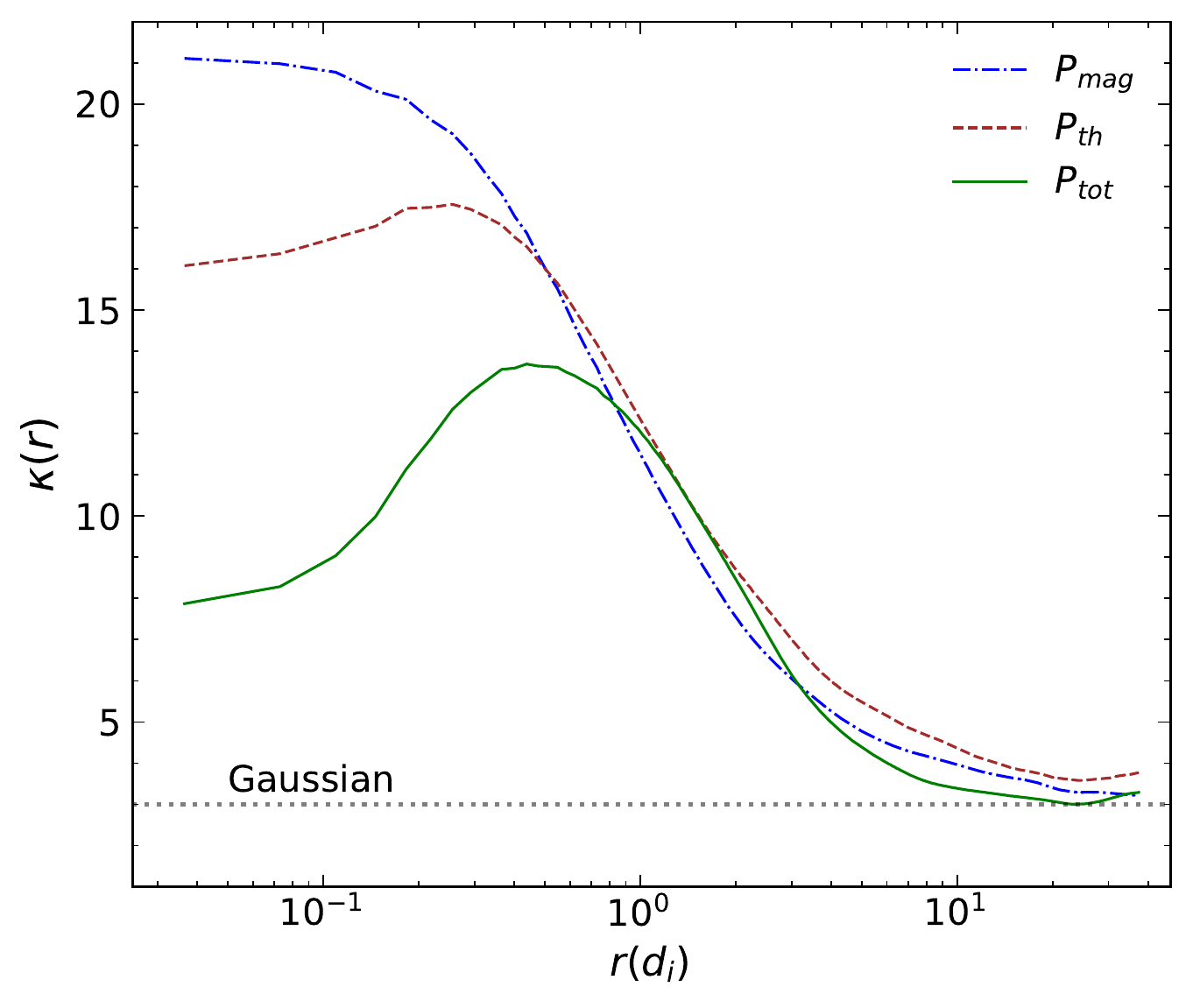}
\caption{\label{fig:sdk}Scale dependent kurtosis $\kappa(\mathbf{r})$ for magnetic (dash-dot), thermal (dashed) and total pressure (solid) as a function of lag. The dotted line is a reference for the kurtosis of a gaussian distribution.}
\end{figure}

In Fig. \ref{fig:sdk} we plot $\kappa(r)$ for magnetic, thermal, and total pressure. It is observed that $\kappa(r)$ is non-gaussian for all pressure forms at the smaller lags $r\leq 1di$. This agrees with the PDF seen in Fig. \ref{fig:pressurepdf}. The scale-dependent kurtosis for the magnetic field pressure peaks at $\approx 0.1d_i$, that for thermal pressure peaks at $\approx 0.3d_i$, while $\kappa(r)$ of total pressure peaks at $\approx 0.5d_i$, approaching a value closer to $3$ for larger lags. This implies that the fluctuations at smaller lags are intermittent, and therefore the original Kolmogorov (1941) similarity hypothesis is not valid and needs some discussion or clarification.

\subsubsection{Structure functions}
In this subsection, we discuss the higher order structure functions for pressure fluctuations as measured by the increments $\delta P$, and compare them with the higher order velocity and magnetic structure functions. The structure function of order $n$ is defined as
\begin{equation}
    D_P^n(r)=\langle \lvert \delta P(r)\rvert^n \rangle. 
\end{equation}

For Kolmogorov 1941 scaling ($K41$ from here on), which ignores intermittency, 
(see section \ref{sec:poisson})
the pressure fluctuation is expected to scale as 
$\delta P \sim (\delta v) ^2$, and 
therefore the $n^{th}$ order pressure structure function $D_P^n(r)\sim (\delta v)^{2n} \sim r^{2n/3}$. However, this idealization is not expected to be valid for higher-order structure functions because of the non-gaussian behavior as seen in Fig. \ref{fig:pressurepdf}. 
This section explores the lag dependence 
of the higher-order pressure structure 
functions in the kinetic plasmas of interest.

The multi-fractal scaling of pressure fluctuations has not been explored as much as the velocity and magnetic field. Some attempts have been made to include intermittency corrections in the pressure scaling ~\citep{donzis2012some}, but not for kinetic plasma turbulence as far as we are aware. 

\begin{figure}
\centering
\includegraphics[scale=0.52]{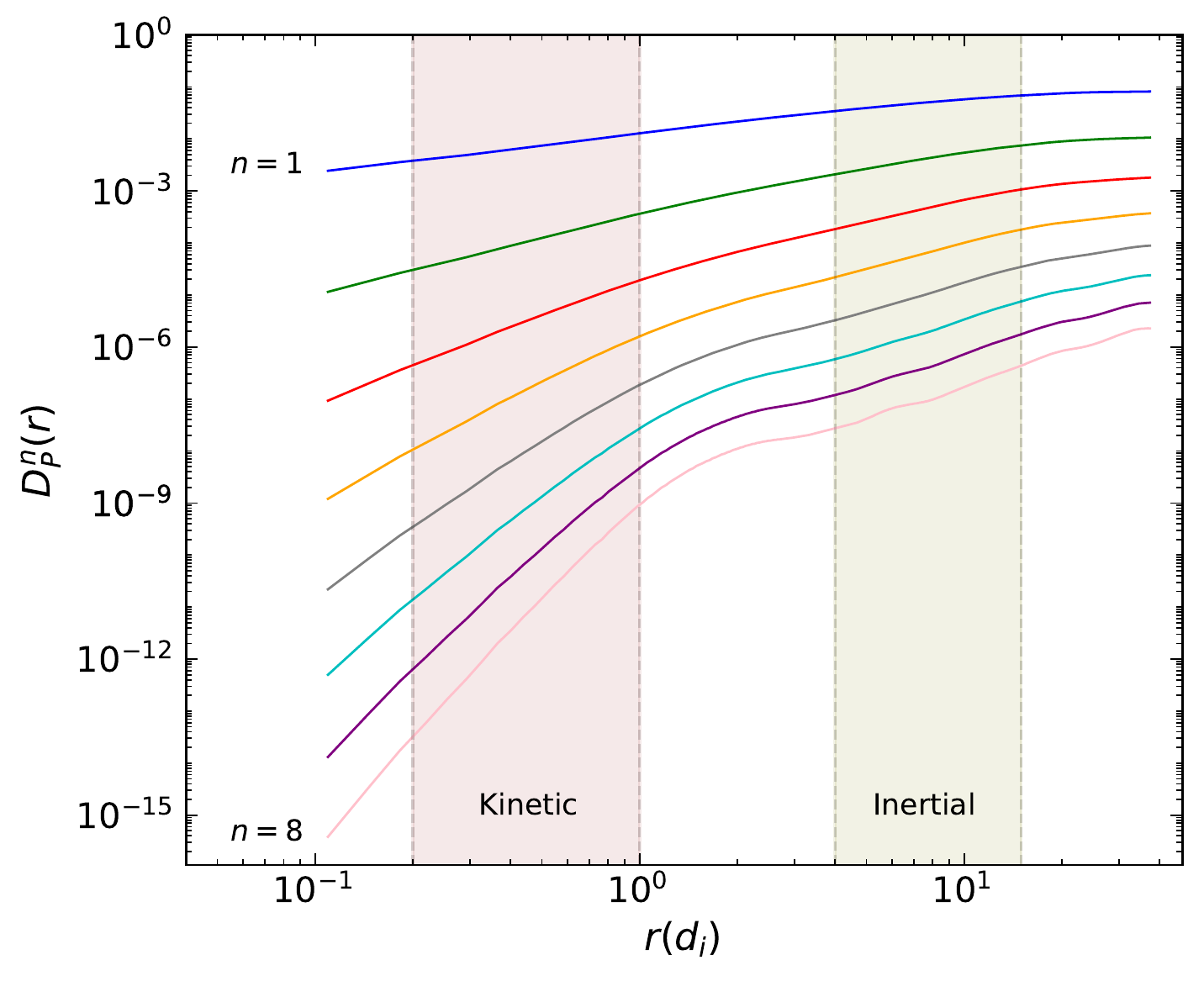}
\caption{\label{fig:higherstf}
Sequence of higher order structure functions for pressure. For uniformity, we assume that the region between the dashed lines at $d_e=0.2d_i$ and $d_i$ denotes the kinetic range (brown rectangle), and the region between $4-15 d_i$ represents the inertial range (olive rectangle) (see~\protect\cite{adhikari2021energy}).}
\end{figure}

For further analysis, first we assume the pressure structure function to take the form
\begin{equation}
    D_P^n(r) \sim r^{\zeta_n},
\end{equation}
where $\zeta_n$ includes the intermittency correction to the $K41$ scaling~\citep{kolmogorov1962refinement,Obukhov62}. We then compute the scaling exponents for our system and compare that with a few intermittency models that are formulated to describe the scaling of velocity structure functions.

One of the first descriptions of intermittency comes from Oboukhov's suggestion that motivated the Kolmogorov refined similarity hypothesis ~\citep{oboukhov1962some,kolmogorov1962refinement}. Specifically Oboukhov reasoned that  patchy dissipation might be accounted for if the velocity fluctuation (i.e., increment) depends on the local value of dissipation 
and exhibits a log-normal distribution. According to this model ($K62$ from here on), the scaling exponent takes the form
\begin{equation}
    \zeta_{n}^{K62}=\frac{n}{3}+\frac{\mu}{18}(3n-n^2),
\end{equation}
where $\mu$ is the intermittency exponent, whose values typically ranges between $0.2-0.3$ ~\citep{van1980reynolds,sreenivasan1993update,davidson2015turbulence}. 

Another successful model used to describe 
intermittency is the She and Leveque model ~\citep{she1994universal} which assumes a log-Poisson model distribution. 
In this model, the exponent has the generalized form
\begin{equation}
    \zeta_{n}=\frac{n}{3}(1-\alpha)+C_0(1-\beta^{n/3})],
    \label{eqn:SL}
\end{equation}
where $C_0=\frac{\alpha}{1-\beta}$ is related to the dimension D of the dissipative eddies through the relation $C_0=3-D$ ~\citep{biskamp2003magnetohydrodynamic}  and therefore called the co-dimension. The parameter $\alpha$ is the scaling exponent, while $\beta$ is a characteristic of the intermittency. For hydrodynamic turbulence, because of the quasi-$1$D structure of thin vortex filaments, the co-dimension $C_0=2$. She and Leveque ~\citep{she1994universal} further assumed that with Kolmogorov scaling, $\alpha=2/3$ and $\beta=2/3$, which yields
\begin{equation}
    \zeta_{n}^{SL}=\frac{n}{9}+2\left[1-\left(\frac{2}{3}\right)^{n/3}\right].
\end{equation}
However, for magnetohydrodynamic turbulence the smallest dissipative structures may be 
taken to be $2$D as we observe from small scale current sheets. For that case, 
$C_0=1$. With the same $x=2/3$, we now have $\beta=1/3$ that yields the scaling exponents ~\citep{politano1995model,muller2000scaling} of the form
\begin{equation}
    \zeta_{n}^{MHD}=\frac{n}{9}+1-\left(\frac{1}{3}\right)^{n/3}.
\end{equation}

\begin{figure}
\centering
\includegraphics[scale=0.55]{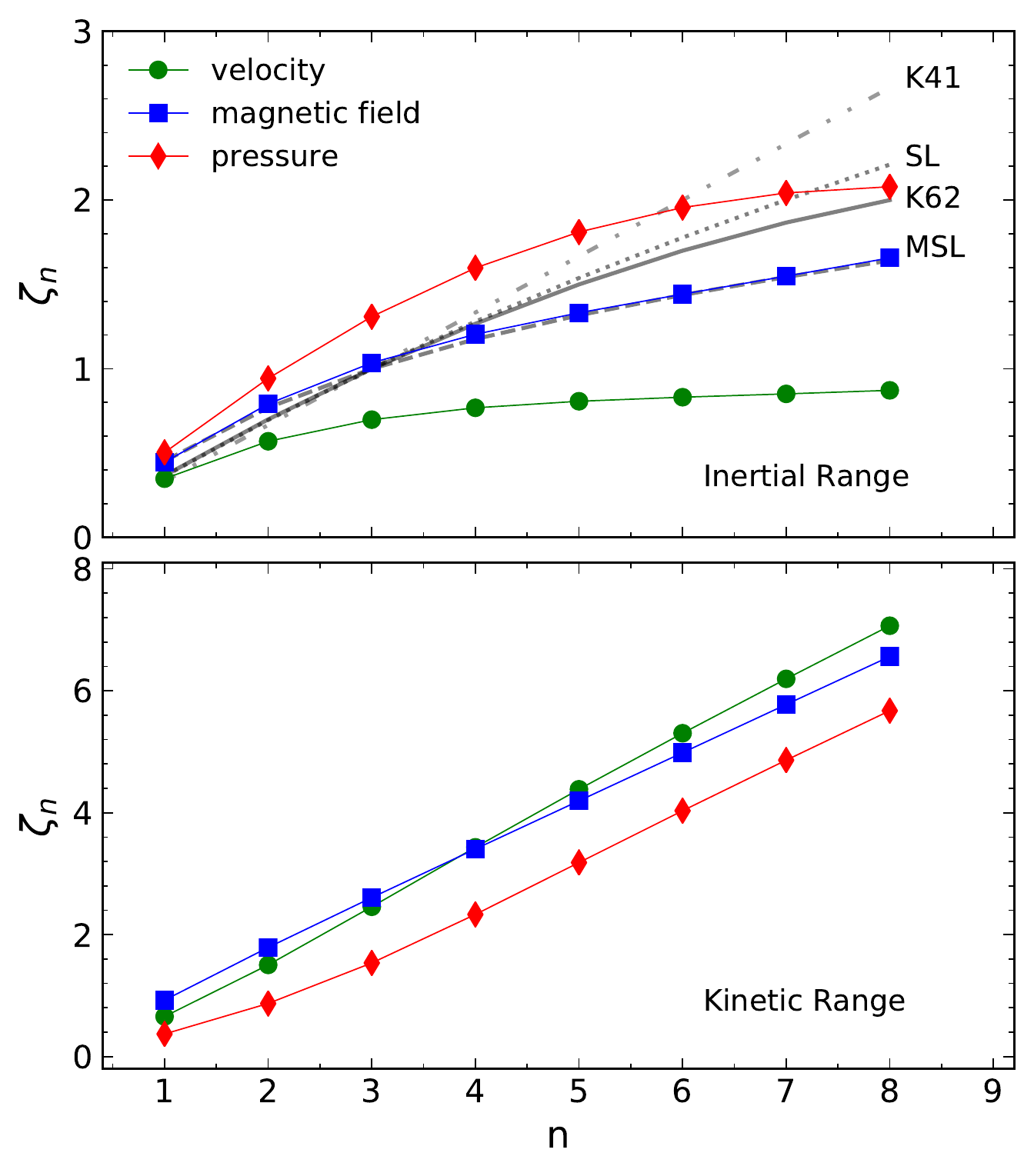}
\caption{\label{fig:higstf_int}Power index for different orders $n$ of the magnetic (blue squares), velocity (green circles), and pressure (red diamond) structure functions in the inertial (top), and kinetic (bottom) range as defined by the shaded rectangles in Fig. \ref{fig:higherstf}. The scaling exponents predicted by different models in the inertial range are drawn for reference (see text for details). In the kinetic range, the scaling exponents follow a linear trend with a slope of $0.927$ (velocity), $0.801$ (magnetic field), and $0.779$ (pressure).}
\end{figure}

Next, in Fig.~\ref{fig:higherstf} we show the higher-order structure functions $D_P^n$ versus lag $r$ up to the eighth order. Two vertical lines on the left denote the kinetic range (between $d_e=0.2d_i$ and $1d_i$), while two vertical lines on the right denote the inertial range (between $4 d_i$ and $15 d_i$) \citep{adhikari2021energy}. The magnitude of the higher-order structure functions is smaller than the lower-order structure functions. The structure functions are observed to be steeper in the kinetic range compared to the inertial range. The steepness of the kinetic range increases sharply with higher-order structure function compared to the inertial range because the fluctuations at those lag scales are much smaller than those in the inertial range. Beyond the inertial range, the structure functions display a similar slope. Next, we calculate the power exponent of these structure functions. In Fig.~\ref{fig:higstf_int} we plot the scaling exponent of these structure functions as a function of the order $n$, across the inertial (top) and kinetic (bottom) range and compare it with those from the magnetic and velocity structure functions. The exponents are obtained using power-law fits to the structure function curve at both the inertial and kinetic ranges.
It is instructive to compare the present results with similar findings presented for kinetic plasma turbulence by \citep{WanEA16} 
and for the density in compressible MHD turbulence by 
\citep{YangEA17-mhd}.

The scaling exponents for the structure function follow the order velocity $<$ magnetic field $<$ pressure, with the ratio between pressure and velocity structure function $\sim 2$. The value of $\zeta_n$ for the velocity/magnetic and pressure structure functions decrease more and more from the conventional (K$41$) value of $n/3$ and  $2n/3$ respectively as $n$ increases. Interestingly, the exponent saturates to a value close to $1$ for the velocity structure function. 

We also plot the K$41$, SL, and K$62$ prediction for comparison. Clearly, none of these models 
fully describe the exponents obtained from our simulations. However, interestingly, the exponent for the magnetic structure functions is in good agreement with the generalized She Leveque exponent (Eqn.~\ref{eqn:SL}) with scaling exponent $x=3/4$, and $\beta=1/4$, such that the co-dimension $C_0=1$. This curve is denoted by MSL (modified SL) in Fig.~\ref{fig:higstf_int}.

In the kinetic range (bottom panel Fig.~\ref{fig:higstf_int}), the exponents for all the structure functions exhibits a linear relation with the order of the structure function, contrary to the inertial range behavior. The values of the exponents for lower-order $(n \lesssim 6)$ magnetic and velocity structure functions are similar. However, for $n > 6$ the difference in $\zeta_n$ starts to increase and diverges with increasing $n$ (not shown). On the contrary, the power exponent for the pressure structure functions is smaller in the kinetic range suggesting that the previous description in the inertial range is not valid in the kinetic range. 
More studies emphasizing the kinetic regime of the structure functions will certainly shed some light on our understanding of the structure functions. 

Recent works on intermittency in the magnetic field 
provide guidance in this issue, while it remains clear that some subtle 
distinctions may be needed. 
For example, \cite{LeonardisEA13}
found multifractal scaling of the magnetic field at subproton scales in simulations of kinetic reconnection, but also 
found monofractal gaussian scaling 
at scales less than $d_i$ \citep{LeonardisEA16}. Other kinetic simulations of 
collisonless plasma turbulence supported the finding that turbulence at proton scales and smaller is self-similar, i.e., monofractal
\citep{WanEA16}.
Laboratory findings (e.g., \citep{schaffner2015multifractal})
and Parker Solar Probe data 
\citep{chhiber2021subproton}
also find 
monofractal behavior
at sub-proton scales. Here we have found very similar results for 
intermittency in the total pressure. 
First we have found that the 
pressure fluctuations are indeed intermittent. 
Second, the finding of of multifractal scaling at scales larger than $d_i$, and monofractal scaling at scales less than $d_i$ is in accordance with characterization of magnetic field intermittency in a number of prior studies. 

\section{Conclusions}\label{sec:conc}

Pressure balance has long been an important element in space plasma physics \citep{BurlagaJGR1990}.
The traditional viewpoint adopts explicitly an assumption 
that fluctuations are small, 
leading to a menu of linear wave solutions and small amplitude 
pressure fluctuations that 
emerge as special static solutions of the MHD equations~\citep{cowling1976magnetohydrodynamics}.
Here we 
have employed  particle-in-cell simulations of kinetic turbulence starting from conditions that give rise to strong 
MHD scale fluctuations and subsequently the formation of 
strong spatial intermittency, coherent structures, and 
nonuniform dissipation. 

In this paper, we emphasize 
the properties of pressure fluctuations that are of inherently nonlinear and 
stochastic nature, and therefore the characteristics
we describe are attributable to turbulence and not a collection of static or wavelike structures. 

The key conclusions of this paper are listed below:
\begin{enumerate}
    \item Poisson's equation for pressure (Eqn.~\ref{eqn:pressurefinal}) in terms of Els\"asser's variable was presented. This equation explains the quasi-pressure balance obtained in turbulent systems. Similarly, Poisson's equation for the pressure fluctuation is formulated in terms of Els\"asser shear (Eqn.~\ref{sheareqnfinal}). The form of the equation is the same as it would be for velocity replaced by Els\"{a}sser variables.
    \item In a PIC simulation of decaying turbulence, as turbulence is fully developed, the volume averaged change in the thermal pressure behaves opposite to that of the magnetic pressure keeping the total pressure nearly constant as seen in Fig. \ref{fig:deltaP}. A 2D view of the thermal (ion plus electron) and magnetic pressure indicates the suggested signatures of anticorrelation between the magnetic and the thermal pressure locally, and results in the global anticorrelation (Fig. \ref{fig:pressure2D}). 
    \item The omnidirectional pressure spectrum has a characteristic slope of $-7/3$ (Fig. \ref{fig:pstf_spectra}) in the inertial range while the magnetic and thermal pressure each have a slope of $-5/3$, a less steep power spectrum than the total pressure spectrum. The significant cross contribution between the two pressure terms (magnetic and thermal) is a result of the anti-correlation between the two.
    \item The intermittent behavior of pressure fluctuations is discussed 
    and contrasted to the intermittency observed in the velocity and magnetic fields. The ratio of the scaling exponents for the higher-order pressure structure function to that of velocity structure function in the inertial range is $\sim 2$. None of the previously described intermittency models fit the scaling exponent obtained from our simulations;
    resolution of this issue
    requires detailed study that is outside the scope of the present paper. The exponents for velocity structure functions are much lower than the predicted value for higher orders $n$ and saturate close to $1$. The scaling exponent for magnetic structure functions does agree with the MHD description of the SL model but with a different scaling exponent. Further, both monofractal and multifractal scaling exponents are observed in the pressure, suggesting that traces of the intermittent behavior are also observed in pressure fluctuations.
 \end{enumerate}
    
This study sheds light on the contribution of thermal and magnetic pressure to the total pressure using pressure spectra and higher-order statistics. One can use these results to characterize the 
fluctuations of mass density
in many astrophysical systems, exploring the contributions coming from the thermal and magnetic pressure in the system~\citep{lau2013weighing}. Therefore, studies of pressure fluctuations in astrophysical systems can only lead to a better understanding of the system.

Finally it is not yet clear if the findings presented here will extrapolate to all the cases of two or three dimensions, 
including the expected 
anisotropy of magnetic fields \citep{OughtonEA94}. Given the great variation in physical processes that occur in widely different parameter regimes of three dimensional plasma turbulence, it is also not clear if the conclusions present here will prove to be universal across all attainable 
parameters. Future studies on these extensions are therefore warranted.

\section*{Acknowledgements}
S.~A. acknowledges helpful discussions with Francesco Pecora. All the authors would like to acknowledge the high-performance computing support from Cheyenne \citep{Cheyenne18} provided by NCAR's Computational and Information Systems Laboratory, sponsored by the NSF. We also thank NERSC resources, a U.S. DOE Office of Science User Facility operated under Contract No. DE-AC02-05CH11231. S.~A, and M.~A.~S. acknowledge support from NASA LWS 80NSSC20K0198. P.~A.~C gratefully acknowledges the hospitality from the University of Delaware and the Bartol Research Institute during his sabbatical visit, and acknowledges support from NASA Grant No. 80NSSC19M0146,  and the Bartol Research Institute. S.~A. and P.~A.~C. are also supported by DOE grant DE-SC0020294. W.~H.~M is supported by NSF DOE grant AGS 2108834 at the University of Delaware, the IMAP project (Princeton subcontract SUB0000317)) and the NASA LWS program FST grant to New Mexico Consortium (subcontract 655-001 to Delaware). 

\section*{Data Availability}
The data that support the findings of this study are openly available in Zenodo~\citep{adhikari_S_2023_dataset}.

\appendix
\section{System Overview}
\begin{figure}
    \begin{center}
        \includegraphics[scale=0.55]{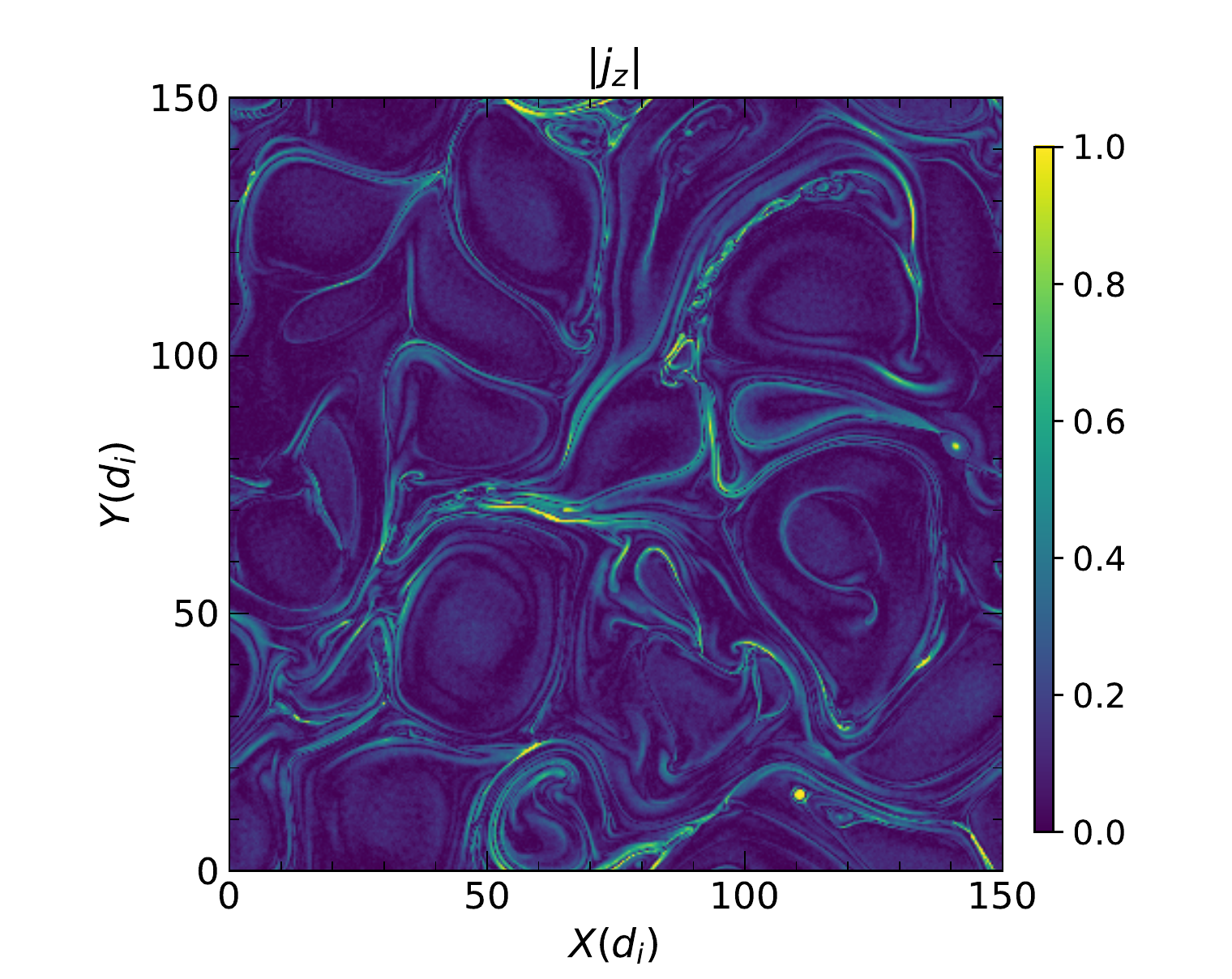}
        \caption{\label{fig:jzmag} Magnitude of the out-of-plane current density $j_z$ at the time of analysis $(t\omega_{ci}=116.5)$. The current sheets are the regions with larger magnitudes of $j_z$.}
    \end{center}
\end{figure}

Fig.~\ref{fig:jzmag} shows the magnitude of the out-of-plane current ($j_z$) in the system at the time of analysis.



\bibliographystyle{mnras}
\bibliography{Pressure,ag,hl,mp,qz,refs_WHM} 




\appendix


\bsp	
\label{lastpage}
\end{document}